\documentclass[12pt]{iopart}

\usepackage{graphicx}
\usepackage{dcolumn}
\usepackage{bm}
\usepackage{cite}
\usepackage{import}
\usepackage{color} 

\newcommand{\fig}[1]   {figure~\ref{#1}} 
\newcommand{\Fig}[1]   {Figure~\ref{#1}} 
\newcommand{\tab}[1]   {Table \ref{#1}}
\newcommand{\Ref}[1] {\cite{#1}}

\newcommand{\D}{$\Delta$ }
\newcommand{\dD}{$\delta\Delta$ }
\newcommand{\alox}{$\textrm{Al}_2\textrm{O}_3$}
\begin{document}

\title[Energy gap measurement of nanostructured thin aluminium films]{Energy gap measurement of nanostructured thin aluminium films for use in single Cooper-pair devices}

\author{N~A~Court,
        ~A~J~Ferguson\footnote[1]{Present address: Microelectronics Research Centre, Department of Physics, Cavendish Laboratory, Cambridge, CB3 0HE, UK.},
        and~R~G~Clark
}%

\address{Australian Research Council Centre of Excellence for Quantum Computer Technology, University of New South Wales, Sydney, NSW 2052, Australia}
\ead{ncourt@phys.unsw.edu.au}
\begin{abstract}
Within the context of superconducting gap engineering, Al-\alox-Al tunnel junctions have been used to study the variation in superconducting  gap, $\Delta$, with film thickness. Films of thickness 5, 7, 10 and 30 nm were used to form the small area superconductor-insulator-superconductor (SIS) tunnel junctions. In agreement with previous measurements we have observed an increase in the superconducting energy gap of aluminium with a decrease in film thickness. In addition, we find grain size in small area films with thickness \textbf{$\geq$} 10 nm has no appreciable effect on energy gap. Finally, we utilize 7 and 30 nm films in a single Cooper-pair transistor, and observe the modification of the finite bias transport processes due to the engineered gap profile. 
\end{abstract}

\submitto{\SUST}
\maketitle
\section{Introduction}

By contacting superconducting materials with different gap energies it is possible to modify the energetics of quasiparticle states in different regions of a superconducting nanostructure. This allows quasiparticles to be either confined in or excluded from certain parts of a device. This principle of quasiparticle gap engineering is used in photon detection where it is desirable to confine excess quasiparticles resulting from a photon absorption event \cite{gaidis_ieee_93,goldie_prl_90,wilson_prl_01}. In addition, the interplay between Coulomb blockade and an engineered gap profile can be used to suppress unwanted quasiparticle tunnelling (QP poisoning) in single Cooper-pair transistors (SCPTs) and Cooper-pair boxes (CPBs) \cite{aumentado_prl_04,yamamoto_apl_06,ferguson_prl_06}.

Methods of gap engineering include the use of different superconducting materials such as aluminium and tantalum\cite{gaidis_ieee_93, wilson_prl_01}, superconductor-normal metal bilayers \cite{goldie_prl_90}, oxygen-doped aluminium \cite{chi_prb_79,aumentado_prl_04} and more recently different thickness aluminium films \cite{yamamoto_apl_06,ferguson_prl_06,ferguson_prl_06_spin}.
This work utilizes the latter method in which changes in the superconducting gap of more than 50$\%$ may be achieved \cite{chubov_jetp_69}.
Early experiments on thin aluminium films observed the enhancement of the superconducting gap and transition temperatures well above bulk values. This enhancement has been attributed to different types of disorder found in thin film structures such as grain size, average lattice constant and the presence of an oxide layer surrounding each grain \cite{cherney_cjp_69,townsend_prb_72,cohen_pr_68,pettit_prb_76}, as well as the film thickness itself\cite{cherney_cjp_69,meservey_jap_71,chubov_jetp_69}. 

Controlling the superconducting gap profile using different thickness aluminium films is a useful concept. In particular, there is no need for introduction of different materials, which can complicate the evaporation process. However, the use of films as thin as 5 nm in Coulomb blockade devices such as the SCPT \cite{ferguson_prl_06_spin} involves evaporation onto a liquid nitrogen cooled stage and requires reliable fabrication techniques. It is also important to identify any unusual characteristics due to added disorder in such films such as the presence of multiple superconducting gaps.

Aumentado \etal \cite{aumentado_prl_04} have recently used oxygen doping to investigate the parity of the supercurrent in a gap engineered SCPT. The parity (whether the supercurrent occurs at odd or even integer gate charge) was found to be strongly influenced by the difference in gap energies of the island and leads $\delta\Delta = \Delta_i-\Delta_l$. Their explanation of this behaviour involves the tunnelling of thermal, or non-equilibrium, quasiparticles from the leads onto the device island. Essentially $\delta\Delta$ is the energy cost of transferring a quasiparticle from the leads to the island, and consequently, it alters the corresponding quasiparticle tunnel rates and occupation probability. For the case of positive $\delta\Delta$ the quasiparticle tunnel rates can be suppressed, and the subsequent use of radio-frequency techniques have enabled several experiments in which quasiparticle tunnelling rates were studied on microsecond timescales \cite{naaman_prb_06, ferguson_prl_06}. The ability to create well known $\delta\Delta$ in single Cooper-pair devices is the main motivation behind this study.

Here we present measurements of small area ($\leq$ 100$\times$100 nm$^2$) aluminium superconductor-insulator-superconductor (SIS) junctions. From the maxima in differential conductance measurements we infer the peak density of states, and thus determine superconducting gap energies, $\Delta$, of d = 5, 7, 10 and 30 nm aluminium films \cite{tinkham}. We focus on single thickness film (S1-I-S1) junctions for more reliability in the determination of gap energies. We also investigate films with different grain sizes, by changing evaporation temperatures, to ascertain if there is any appreciable effect on energy gap. Finally, we briefly present measurements on a gap engineered SCPT (30 nm - 7 nm - 30 nm). We discuss both the 2$e$-supercurrent and the finite bias resonances.

\section{Thin-film Junction Fabrication}
\begin{figure}
\begin{center}
\begin{tabular}{ll}
	\textbf{(a)}\\
	\multicolumn{2}{l}{\includegraphics[width=8.25cm]{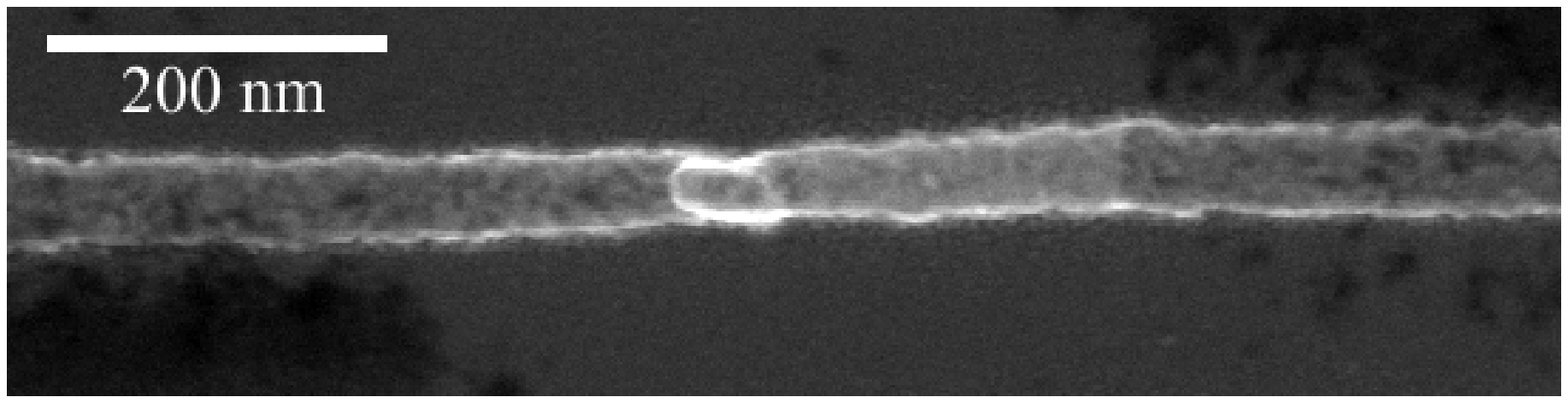}}\\
	\textbf{(b)} & \textbf{(c)} \\
	\includegraphics[width=3.92cm]{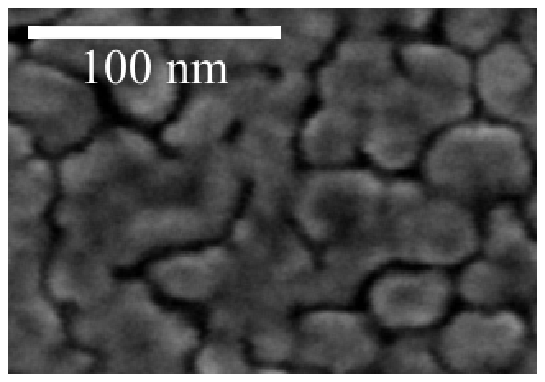}
&	\includegraphics[width=3.92cm]{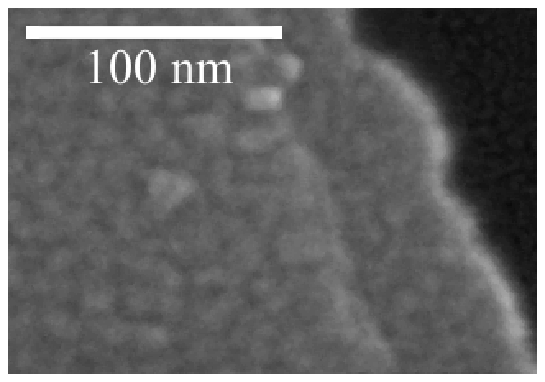}
\end{tabular}
\caption{Thin film Al-\alox-Al junctions were fabricated using a Dolan bridge resist structure and oxidized in-situ. (a) shows a scanning electron micrograph (SEM) of a typical device structure, with a junction size of $\sim $50$\times$75 nm$^2$. SEM images of the grain structure of an aluminium thin film (thickness 10 nm) in (b) a room temperature evaporated device T $\sim $ 293 K and (c) device evaporated onto a liquid nitrogen cooled stage at T $\sim$ 173 K . }\label{semsis}
\end{center}
\end{figure}

\Fig{semsis} shows a typical test sample with a junction area of $\sim$ 50$\times$75 nm$^2$. A number of devices were measured with resistances ranging between $\sim$ 20 $\textrm{k}\Omega - $2  $\textrm{M}\Omega$. 
Devices were fabricated on a high resistivity (5 k$\Omega\cdot$cm) silicon substrate with 200 nm of thermally grown silicon dioxide on the surface. To enable sufficient contact to the thinner films ($\leq$ 10 nm) a single layer of PMMA was patterned to define thin overlap areas ($\sim$ 15$ - $20 nm Au/Ti) between the optically defined Au/Ti bond pads and the EBL-defined device structure. For the SIS structures a Dolan bridge resist structure\cite{dolan_apl_77} was formed from a UV-sensitized 33\% copolymer of PMMA/MA and a high resolution PMMA. After exposure, the devices were developed in a mixture of MIBK:IPA 1:3 for 60 s. 

In order to achieve consistently continuous films down to thickness's of less than 10 nm the films were evaporated onto a cooled substrate \cite{meservey_jap_71}. With this method we have achieved electrically continuous films of 5 nm in thickness \cite{ferguson_prl_06_spin}. Films were evaporated in a commercial Edwards evaporator with a custom oxide chamber and rotational stage. Liquid nitrogen could be introduced into the chamber via a modified feed through into the top of the oxide chamber, making contact with the stage onto which the substrate was mounted. 

The temperature of the stage was monitored via a thermocouple in contact with the back of the stage, giving an indication of the substrate temperature during evaporation. Cooled evaporations were performed at $T \simeq$ 173 K whilst room temperature evaporations were at $T \simeq$ 293 K. The variation in temperature during evaporation was $\leq$ 5 K. The insulating barrier was grown in-situ with the first aluminium layer exposed to 4.5 $\times$ 10$^{-2}$ mbar O$_2$ for 4 min. We found that a steady evaporation rate is particularly crucial in achieving continuous films of 10 nm, for a room temperature evaporation, and 5 nm, for a cold evaporation. The aluminium was evaporated at a rate of 0.1 nm$\cdot$s$^{-1}$. 

The grain structure of the films depends strongly on evaporation temperature. A scanning electron micrograph of the grain structure of a 10 nm film evaporated at 293 K is shown in \fig{semsis}(b). Large structure is observed with grains of up to 40 nm in diameter seen. In contrast grain structure in the 10 nm film evaporated at 173 K (\fig{semsis}(c)) is significantly smaller with the largest grain size $<$ 10 nm in diameter.

\section{Results and Discussion}

\begin{figure}
\begin{center}
\includegraphics[width=8cm]{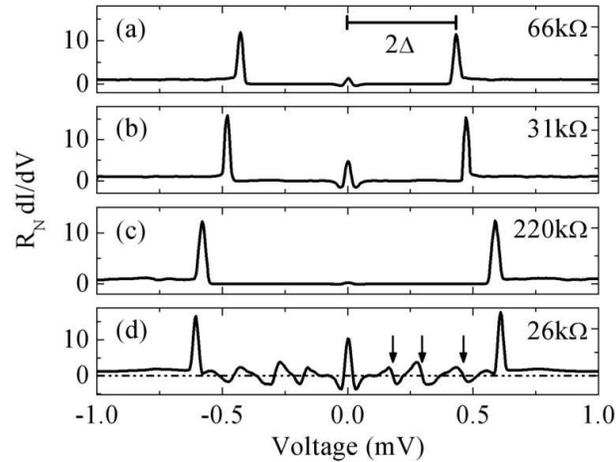}
\caption{Normalized differential conductance as a function of applied bias voltage for 4 different thickness films. 2$\Delta$ was determined from the peak in differential conductance. The 4.2 K resistance of each device is also shown. (a) d = 30 nm, $2\Delta = 430\mu\textrm{eV}$ (b) d = 10 nm, $2\Delta = 476\mu\textrm{eV}$ (c) d = 7 nm, $2\Delta = 584\mu\textrm{eV}$ and (d) d = 5 nm, $2\Delta = 608\mu\textrm{eV}$. (d) The dash-dot line indicates $dI/dV =$ 0 and the points at which $dI/dV$ crosses from $>$ 0 to $<$ 0 are indicated by arrows. }\label{dIdV}
\end{center}
\end{figure}

Measurements were performed in a dilution refrigerator with a base temperature of approximately 100 mK. The differential conductance ($dI/dV$) was measured with a standard low-frequency ac-lock-in technique with a modulation amplitude of 10 $\mu\textrm{V}$. \Fig{dIdV} shows plots of typical differential conductance vs applied bias voltage for four different thickness films. 

Maxima corresponding to the peak in density of states at the gap edges allows accurate determination of 2$\Delta$ \cite{giaever_prl_60}. There is a clear dependence on the observed 2\D as a function of film thickness ranging from 2\D= 430$\mu\textrm{eV}$ for the 30 nm film (see \fig{dIdV}(a)), to an increase of almost 50$\%$ for 5 nm films with 2\D = 608$\mu\textrm{eV}$ (see \fig{dIdV}(d)). The differential conductance has been rescaled to a normalized resistance due to differences in junction area as indicated by the normal state resistances given in \fig{dIdV}. 
It is of interest to note that only singular peaks occur in the quasiparticle density of states. This is opposed to multiple gap structures that are sometimes observed in disordered films \cite{ruggiero_prb_87}.

Subgap structure in the 5 nm devices (see \fig{dIdV}(d)) is more
complicated, perhaps due to the lower junction resistance (R =
26 k$\Omega$) of this sample. We see negative differential resistance regions ($dI/dV <$ 0) which lead to peaks in the integrated $dI/dV$ (not-shown). Similar behaviour was also seen in a second device (R = 21 k$\Omega$). At present we cannot attribute this behaviour to specific transport processes. However, we note that multiple Andreev reflection is unlikely to be the origin since the positions do not coincide with integer multiples of 2$\Delta/n$ \cite{klapwijk_physica_82}.

\begin{figure}
\begin{center}
\includegraphics[width=8cm]{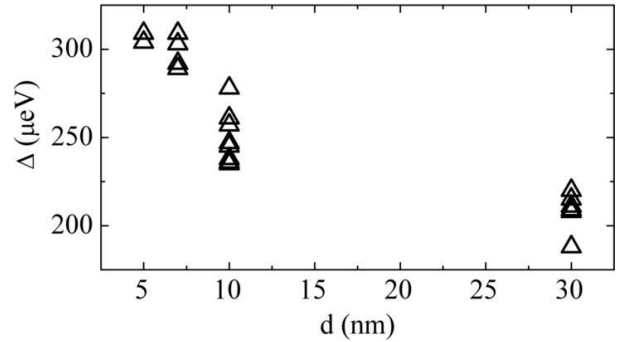}
\caption{Observed superconducting energy gap \D of aluminium as a function of film thickness. Each point represents a single junction whose gap energy was extracted via differential conductance measurements. Films were evaporated onto a substrate in contact with a liquid nitrogen cooled stage at a temperature of $\sim 173$ K with grain structure similar to that seen in \fig{semsis}(c).}\label{thick}
\end{center}
\end{figure}

\begin{table}
\begin{center}
\caption{SIS parameters for cold and room temperature evaporations. Number of samples measured, evaporation temperature, evaluated gap energies and standard deviation for each thickness film.\\ }\label{tablethick}
\begin{tabular}{c@{\hspace*{.05\linewidth}}c@{\hspace*{.05\linewidth}}c@{\hspace*{.05\linewidth}}c@{\hspace*{.05\linewidth}}c@{\hspace*{.02\linewidth}$ \pm $\hspace*{.02\linewidth}}r@{.}l}\hline
\hline
d & Number of &Approx. Junct.& T$_\textrm{evap}$ & \D &\multicolumn{2}{r}{$\sigma$\D}\\
(nm) & Samples & area (nm)$^2$ & (K) & \multicolumn{3}{c}{($\mu$eV)}\smallskip\\
\hline
\\ 

\smallskip
5 & 2 &$100\times100$& 173 & 307 & 3&5 \\ 
\smallskip
7 & 4 &$30\times30$& 173 & 298 & 9&4 \\ 
\smallskip
10 & 8 &$50\times70$& 173 & 250 & 14&9 \\ 
\smallskip
10 & 2 &$50\times70$& 293 & 236 & 3&5 \\ 
\smallskip
30 & 6 &$50\times70$& 173 & 209 & 11&0 \\ 
\smallskip
30 & 1 &$50\times50$& 293 & \multicolumn{2}{l}{208}\\
\hline
\hline
\end{tabular}
\end{center}
\end{table}

\Fig{thick} shows the distribution of the measured energy gap \D as a function of film thickness, d. We observe a increase in \D as the thickness of the film is decreased. The magnitude of this increase is consistent with other studies which infer gap energies from the critical temperature of the films \cite{cherney_cjp_69,meservey_jap_71,townsend_prb_72}. 
From the standard distribution of gap energies taken at each thickness we see a spread of up to 15$\mu$eV away from average values [\tab{tablethick}]. The magnitude of this variation is likely to be due to changes in conditions (e.g. evaporation rate and substrate temperature) between successive evaporations. We support this by noting that junctions made in the same evaporation (for example the 5 nm and the 10 nm room temperature evaporated films) tend to have similar characteristics. The gap energies agree with those obtained in \cite{yamamoto_apl_06} and Table 1 indicates the variation in $\Delta$ that can be easily achieved both by room temperature and cooled substrate evaporations.

Films with larger grain structure [evaporated at 293 K, see \tab{tablethick}] have peaks at $eV =$ 2$\Delta$ which fall within the standard deviation of energies for both the 10 and 30 nm films evaporated at 173 K. Consequently, we see no indication that the grain size has a strong influence on energy gap. Our results agree with the conclusions of previous investigations \cite{cherney_cjp_69,cohen_pr_68} that grain size alone does not account for the enhancement in energy gaps. 

To review, we have presented differential conductance measurements of small area SIS junction which show an enhancement of superconducting gap with decreasing thickness films. A variation of up to 15$\mu$eV from average values for each thickness is accounted for due to variability in evaporation conditions. We see no significant dependence of gap energies on grain size for thicker films.

\section{Thin-Films in the Single Cooper-Pair Transistor}

\begin{figure}
\begin{center}
\begin{tabular}{ll}
(a) & (b) \\
\begin{minipage}[h]{4cm}
\includegraphics[width=3.5cm]{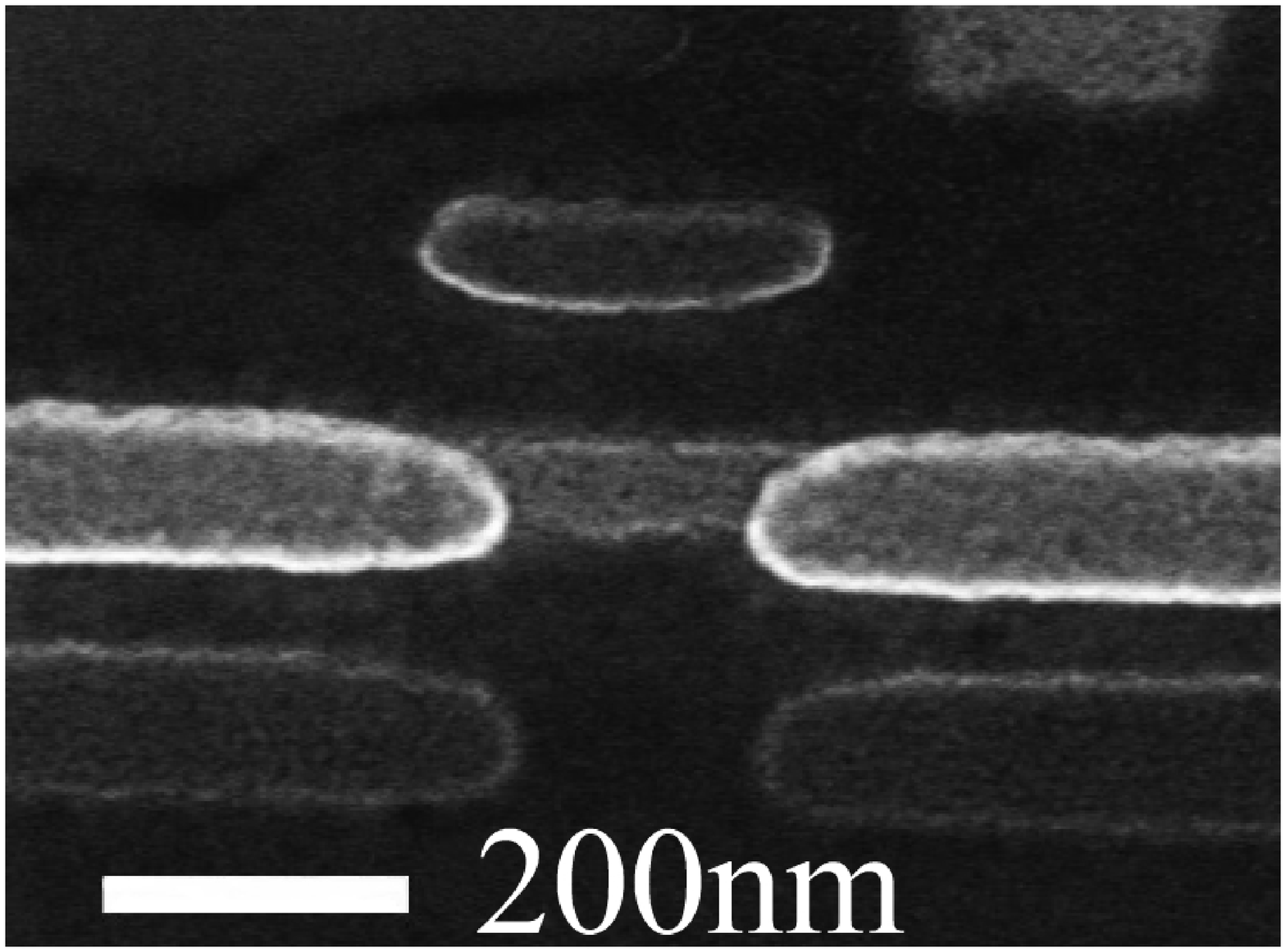}
\end{minipage}
&
\begin{minipage}[h]{4cm}
\includegraphics[width=3.5cm]{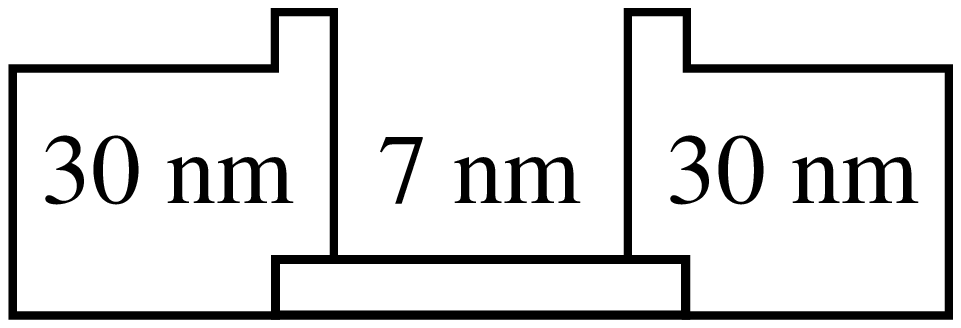}
\\
\end{minipage}\\
\end{tabular}

\begin{tabular}{l}
(c)\\
\begin{minipage}{8cm}
\includegraphics[width=8.0cm]{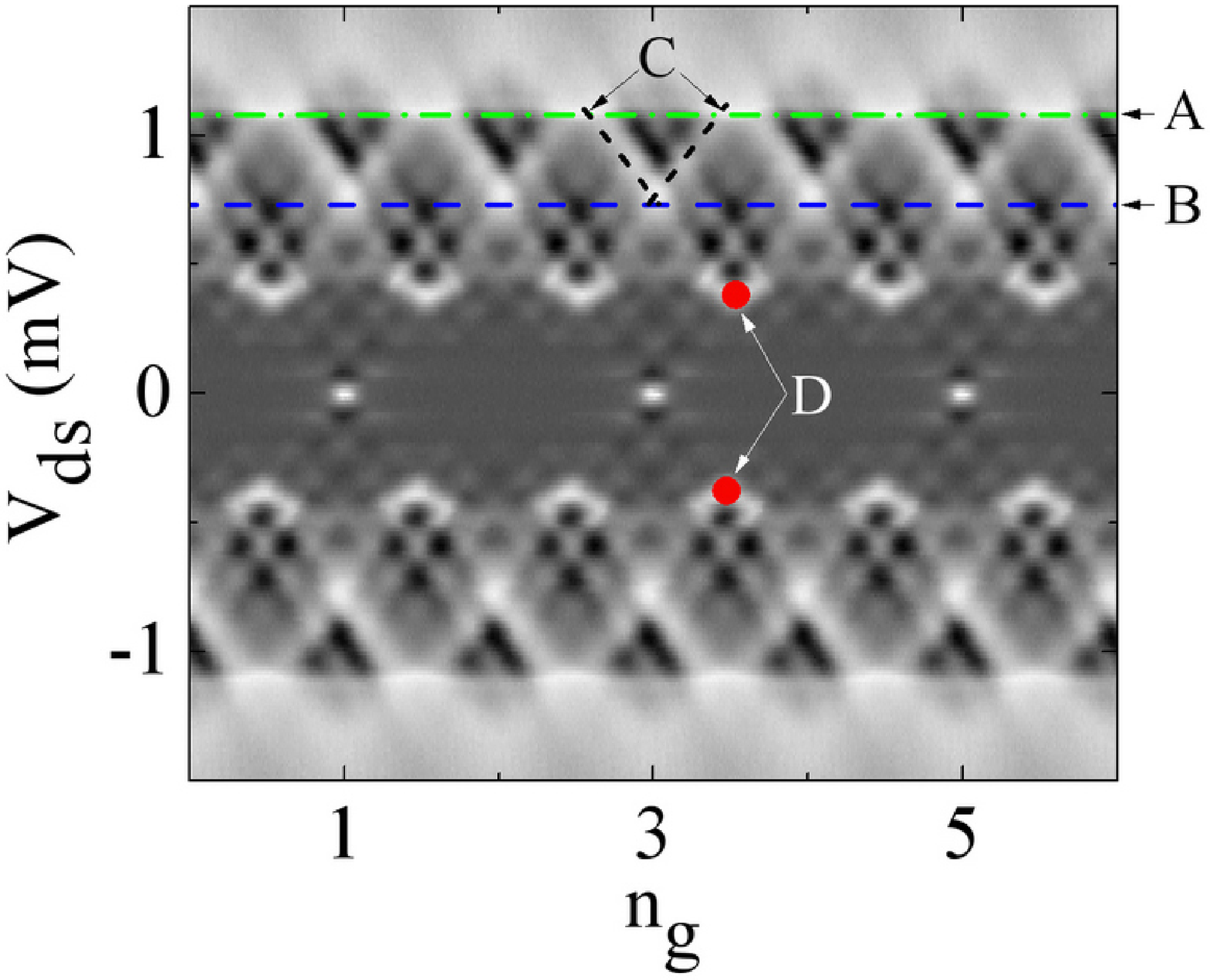}
\end{minipage}\\
\end{tabular}
\caption{We utilize the our knowledge of energy gap and film thickness to design a SCPT with \dD$\sim$ 90$\mu$eV using a 7 nm film ($\Delta \sim$ 298$\mu$eV) for the island and 30 nm film ($\Delta \sim$ 209$\mu$eV) for the leads.
(a) Scanning electron micrograph of a device similar to that measured showing the contrast difference between the 7 nm island and the 30 nm leads. (b) Schematic profile of the change in thickness across the SCPT. (c) Observed superconducting Coulomb diamonds of a SCPT device with a 7 nm island and 30 nm leads. 2$e$-periodic supercurrent peaks can be clearly seen at zero bias, while at finite bias resonances corresponding to $e$-periodic transport are seen (indicated in figure). \textbf{A} corresponds to $eV_{ds} =$ 2$\Delta_i+$2$\Delta_l$, \textbf{B} corresponds to the lower threshold for the JQP cycle occurring at $eV_{ds} = E_C+\Delta_i+\Delta_l$, \textbf{C} shows the condition for the JQP cycle, and \textbf{D} shows the DJQP occurring at $eV_{ds} = \pm$ 2$E_C$.}\label{diamonds}
\end{center}
\end{figure}

The aim of this work is to be able to design superconducting single Cooper-pair devices with well-controlled superconducting gap profile, hence in this section we briefly describe electrical transport measurements on such a sample. We made SCPTs using the same fabrication procedure and junction areas as the SIS tunnel junctions (see \fig{diamonds}(a)). The measurements were performed at milliKelvin temperature in a dilution refrigerator by a radio-frequency reflectometry technique \cite{schoelkopf_sci_1998}. The SCPT was embedded in a resonant LC tank circuit and the reflection coefficient of an incident radio-frequency signal at the circuit resonance frequency is related to the differential conductance of the SCPT  \cite{naaman_prb_06,ferguson_prl_06}. We use the same rf-setup as described in \Ref{ferguson_prl_06}.

From the SIS results presented above we are able to estimate \dD $= \Delta_i-\Delta_l \sim$ 90$\mu$eV for a 30 nm - 7 nm - 30 nm SCPT (\fig{diamonds}(b)).
The 4.2 K resistance of this device was 54 k$\Omega$ and the charging energy $E_C = e^2/$2$C_\Sigma =$ 190$\mu$eV, as determined from normal-state Coulomb diamonds measured at $B =$ 3 T. Estimating the Josephson energy per junction from the 4.2 K resistance and the Ambegoakar-Baratoff relation ($E_J \sim \frac{h\Delta_{i}\Delta_{l}}{4(\Delta_{i}+\Delta_{l})e^2R}$) we find $E_J =$ 30$\mu$eV.

In the Coulomb diamonds (see \fig{diamonds}(c)) we see peaks in the reflected power corresponding to 2$e$-periodic supercurrent in $n_g$ at zero bias. Supercurrent peaks are not observed in the case of a device without gap engineering and their presence is due to a reduced quasiparticle occupation probability on the island. The appearance of the supercurrent due to the quasiparticle barrier on the island is consistent with previous studies \cite{aumentado_prl_04,yamamoto_apl_06}.

Coulomb blockade of quasiparticle tunnelling occurs for $eV_{ds}>$ 2$\Delta_i+$2$\Delta_l$, and from the threshold of this process we can estimate the superconducting gaps of the leads and island. In this device 2$\Delta_i+$2$\Delta_l =$ 1.08meV, and taking the $\Delta_{30nm} =$ 209$\mu$eV, then $\Delta_{7nm} =$ 331$\mu$eV. We note that this is larger than expected for a 7 nm film and is again likely due to be caused by a variation in evaporation conditions. Qualitatively we notice the presence of a large quasiparticle co-tunnelling current in the gap engineered devices.

At finite bias, but for $eV_{ds}<$ 2$\Delta_i+$2$\Delta_l$, there are $e$-periodic features corresponding to the sequential tunnelling of both Cooper-pairs and quasiparticles (\fig{diamonds}(c)). We observe $e$-periodic peaks related to resonant Josephson quasiparticle (JQP) and double Josephson quasiparticle (DJQP) cycles \cite{fulton_prl_89,maassen_prl_91,nakamura_prb_96}. The JQP cycle consists of the coherent tunnelling of a Cooper-pair through one junction followed by two quasiparticles through the other. The condition for Cooper-pair tunnelling must be satisfied and, additionally, the energetics must permit the subsequent tunnelling of two quasiparticles. This results in $e$-periodic repetition of a pair of crossed-lines in the range $E_C+$2$\Delta<eV_{ds}<$ 3$E_C+$2$\Delta$. In principle, the primary change to the JQP cycle due to the modified gap is that the thresholds change so that the cycle occurs in the range $E_C+\Delta_i+\Delta_l<eV_{ds}<$ 3$E_C+\Delta_i+\Delta_l$. The lower threshold is plotted in \fig{diamonds}(c) with the estimated values for the superconducting gaps and appears to correspond to the start of the JQP cycle. 

In the DJQP cycle, coherent tunnelling of Cooper-pairs between the island and junctions is allowed with quasiparticle tunnelling events determining which junction is on resonance \cite{clerk_prl_2002}. For the cycle to be possible, the quasiparticle events must be permitted to occur but, due to the need to satisfy Cooper-pair tunnelling across both junctions, its position must remain fixed at $eV_{ds} = \pm$ 2$E_C$. The DJQP cycle can be seen in \fig{diamonds}(c). 

Additional transport resonances appear just above the DJQP, and parallel to the JQP. Unlike the previously described processes, these differ between devices and may be due to transport though higher order bands in the SCPT or more complex cycles involving both Cooper-pair and quasiparticle tunnelling.

To summarize, the fabrication of a gap engineered SCPT with a positive $\delta\Delta$ has a number of different effects on the transport processes. The primary difference is the appearance of a supercurrent which is discussed in greater detail in \cite{aumentado_prl_04}. Furthermore, the thresholds for Coulomb blockade of quasiparticle tunnelling and the Josephson quasiparticle resonance are shifted. Explanation of the additional resonant features will require a more detailed study of the energetics perhaps involving knowledge of higher order bands.

\section{Conclusion}

We have observed an increase in the superconducting energy gap of aluminium, as a function of decreasing film thickness in small area SIS junctions. We find that our results are consistent with previous studies and that cold evaporation of thin films produces high quality small area junctions which can be used in superconducting gap engineered devices. Further work needs to be performed to decrease the variability in the measured gap especially for the thinnest films. For SCPT engineered from different thickness films we observe 2$e$-periodic supercurrent and a modification of the finite bias transport processes. In addition, for this technique to be used in quantum nanostructures, such as the Cooper pair box, a measurement of the charge noise in the cold-evaporated film and fluctuator density of the tunnel junctions remains to be undertaken.
 

\ack{ The authors would like to thank T. Duty for helpful discussions and D. Barber and R. P. Starrett for technical support. This work is supported by the Australian Research Council, the Australian Government, and by the US National Security Agency (NSA) and US Army Research Office (ARO) under Contract No. W911NF-04-1-0290. }

\end{document}